\documentstyle[12pt,aaspp4]{article}
\def\pspc {3}
\def\ref{\par \noindent \hang}

\begin{document}
\baselineskip 18.0pt
\hspace*{-0.5cm}
\title { Extended  X-ray emission  in the radio loud
galaxy 3C382}
 
\author {M. Almudena Prieto\footnote{Present address: European
Southern Observatory, D-85748 Garching, Germany}}

\affil{Instituto Astrofisica de Canarias, E38200 La Laguna,
Tenerife, Espa\~na\\
Max-Planck-Institut f\"ur extraterrestriche Physik,
D-85748 Garching, Germany}

\begin{abstract}

ROSAT/HRI observations of the powerful radio-loud galaxy 3C382 reveal
extended X-ray emission associated with the source. 
  On the basis of this new spatial component,
a previous  ROSAT/PSPC  spectral analysis of the source  is
revised.  Allowing for the presence of an additional thermal component
in the PSPC spectrum,
the non-thermal component  is found to be
compatible with the extrapolation of the well defined
3C~382 ---2 - 10 keV--- power-law spectrum into
the soft X-rays.  The thermal --extended-- component would then
account for the 
soft excess emission previously reported for  this source. 
The origin of this thermal component is not clear. 
Its luminosity  compares with that of rich Abell
clusters; yet,  the  galaxy environment in 3C382 appears of moderate
optical richness.
An alternative is that it is due to a massive
extended gaseous atmosphere  sustained  by the  deep gravitational
potential well of 3C382.\\
\end{abstract}


\section{Introduction}

Soft X-ray excess emission above a simple  extrapolation of the hard energy
spectrum  is found in a
considerable number of AGNs, mainly  radio-quiet
sources (see Mushotzky, Done, \& Pounds 1993 and
references therein). There is a growing body of evidence for its
spectral ubiquity below $\sim$2 keV or so in Seyferts (e.g. Pounds et 
al. 1994) as in highly luminous quasars (e.g. Saxton et al. 1993). 
Most  previous studies before ROSAT converge toward the idea that the soft
excess is a rather common feature among  radio-quiet quasars, whereas it is
almost absent in their radio-loud counterparts.
{\em ROSAT}/PSPC data of radio loud sources  have shown that a soft excess 
component is also present in  radio loud sources
 (e.g. Buehler et al. 1995; Prieto 1996; Siebert et al. 1998).

 Possible interpretations  for  the soft X-ray excess in AGNs include 
thermal emission 
from the inner regions of an accretion disk, scattering by highly 
ionized material in its vicinity (Pounds et al. 1986; Ross \& Fabian 1993), 
or thermal emission due to shock-heated gas in the close 
vicinity of the nucleus  (Viegas \& Contini 1994).
The poor spatial resolution of the {\em ROSAT}/PSPC makes difficult  
the  separation 
between possible components of the observed  emission. Indeed, 
 the large     PSPC resolution beam, $\sim$ 25 arcsec at
 1 keV, makes  
plausible that an important part of the observed  emission to be due to 
 an extended gas component surrounding the AGN. 
In the particular case of radio-loud galaxies which are characterized by  large radio sizes, a
 hot surrounding medium becomes a  necessary component   for providing the
working surface for  the radio emission.
In the  analysis of the 3CRR sample by Prieto,  
 a first attempt to   fit the  PSPC spectra of 
sources  with extended emission --mostly in Fanaroff \& Riley (1974)
 type I sources (FRI)-- with a single power-law leaded
 to extreme step spectral index, the reason being  due to  
the dominant contribution of the gaseous medium in which those sources
usually reside.  In the case of   FRII, 
a single power-law fit provided
 a fair representation of the PSPC spectrum but with average
 spectral index about -1.1, and so above the extrapolation of the
 canonical  hard-energy spectrum 
into the soft X-rays.  Clustering of galaxies about FRII sources is
 less common than in FRI, in particular at low redshift; yet,  FRII
could contain  their own extended   gaseous atmosphere which 
may  directly  
translate into a steepening of the PSPC
spectrum. This component  however may prove to be elusive with present
 X-ray instrumentation. 

This paper presents deep {\em ROSAT}/HRI observations of the powerful 
X-ray radio-loud source 3C~382. This is one of the few  nearby 
 broad-line  galaxies ($z= $0.0578) that show extremely bright and broad
permitted lines (FWZI$> 25000 km~s{-1}$; Tadhunter, Perez \& Fosbury
1986) and a strong continuum, with
 with a X-ray luminosity in 0.2-2.4 keV band  of 
$L_{\rm x} \sim 7.10^{44}$ erg~s$^{-1}$ (Prieto 1996),
 and a radio power at 178 
MHz of  $L_{\rm 178MHz} \sim 3.10^{33}$ erg~s$^{-1} Hz^{-1}$ 
(Laing et al. 1983). 

{\em EXOSAT} monitoring  of the source (1983--1985)
tightly constrains the high-energy (above 2 keV)  spectral index of 3C382 to   
$\alpha=-0.7\pm 0.1$ (Ghosh \& Soundararajaperumal 1992). This  is also
confirmed by more recent ASCA data (Wozniak et al. 1998). However, the
ROSAT/PSPC spectral analysis  of the source  shows 
compatible with a 
power-law model  with spectral index $\alpha= -1.2\pm0.3$, and absorbed
by  a column density, $N(H)= 0.78\times 10^{21} cm^{-2}$, that is in agreement with the
Galactic value. Thus, 3C382
shows a  soft excess emission below $\sim$ 2keV (Prieto 1996). 
Independently, the  presence of a soft excess is also inferred from the 
analysis of the EXOSAT (Ghosh \&
Soundararajaperumal), ASCA (Wozniak et al.) and Ginga  (Kaastra et al
1991) data.

Extended soft X-ray  emission associated
with this source  is detected in the ROSAT/HRI data. 
On the basis of that new component a
re-evaluation of the PSPC spectrum  is presented.

Throughout this work $H_{0}$ = 50 km s$^{-1}$ Mpc$^{-1}$. 1 arcsecond 
corresponds to $\sim$ 1.7 kpc at the source.

\section {Analysis of HRI data}

The HRI observations of 3C~382 were conducted in 1996 October and 1997
April (WG900720H and WG900720H-1 datasets respectively)
The corresponding total accepted times were 4514 s and 13310 s,  respectively.
The counts are integrated from 
channels 1 to 8 which enclose most of the  energy accumulated in the 
ROSAT band, yielding for both dataset count rates of $\sim$1 cts s$^{-1}$.

The nominal resolution of the {\em ROSAT}/HRI is  $\sim$ 5--6 arcsec (FWHM).
Residual errors in the ROSAT aspect solution are known  
 to give rise to elongated images (David et al. 1996), 
the shape of the surface brightness  profile dramatically departing from the
expected point response function (PRF). In the case of 
very bright sources,
improvement of the  HRI spatial resolution   becomes feasible  by
using speckle interferometric techniques such as the ``shift-and-add'' method. 
If one constructs images over short time 
intervals, the PRF becomes symmetric and therefore the elongation in 
the image appears as an apparent residual motion of 
the X-ray source in the sky.   
Such  residual motion can be corrected for by de-speckling.

3C~382 is  particularly  suitable for that technique as it shows 
 very bright  
in the {\em ROSAT} band, with $\sim$ 2 cts s$^{-1}$ in 
the PSPC and $\sim1$ cts s$^{-1}$ in the HRI.

The procedure used  follows the same criteria and approach originally 
presented in Schmitt, G\"udel \& Predehl (1994). 
Basically, each  event file  is divided into time
bins of typically 50 s. For each bin,   the apparent X-ray position of the
source (i.e., RA and $\delta$) is determined as a function of the observing 
time.  A  spline function
is fitted to these data points,  all recorded photons being then corrected 
with the appropriate time-dependent correction in RA and $\delta$.

To validate the correction, new 
measurements of the  source centroid are  repeated on the 
corrected event file. The correction is considered as satisfactory
 if the new centroid positions cluster during the time period of the 
observation about   an average constant value. The  uncertainty in the final
source centroid is  about 2.5 arcsec in RA and 1.5 arcsec in $\delta$.

The X-ray spatial analysis   was then 
performed on  the corrected  HRI event files. 
Because of the much higher statistic of the 
 April event file, reliable results from the  
 de-speckle procedure are only found from that dataset. Thus, the   
following  analysis focuses  only on  this dataset.

\section{The extended X-ray component}

An X-ray contour image of 3C~382 obtained after the de-speckle
procedure is shown in Figure 1a.  The image is background-subtracted
and smoothed with a Gaussian filter of FWHM $\sim$12 arcsec. The
background level is estimated from different regions in the image
within the central 5 arcmin.  The emission is dominated by a central
peak component, and a surrounding  halo slightly more
asymmetric towards the North-East  side.
 Some of the morphological features are
also apparent in the shortest exposed 1996 October event file -not
shown.  Overall, the X-ray emission extends out to about 100 arcsec
from the center, $\sim$170 kpc at the distance of the source. 
This is about the  size of the radio structure at 8.3 GHz
(Black et al. 1992). The  slight North-East asymmetry 
 in the X-ray image is  virtually coinciding with  the  
direction of the radio axis  and with that   of
the faint filamentary regions and bright companion galaxy seen in the 
HST/WFPC2 image (Fig. 1b).

The corresponding surface-brightness profile is presented in Figures
2.  For comparison, the new improved {\em ROSAT}/HRI PSF profile by
Predhel (1998) ---currently upgrade in the EXSAS package--- which
includes the effect of the ROSAT mirror scattering and has been fit to
Capella and Sirius, is shown superimposed. To illustrate the effect of
the de-speckle procedure on the data, the surface brightness profile
as derived from the original data is shown is Fig 2a, and that after
de-speckling in Fig. 2b.  The departure from the ROSAT/HRI PRF is
clearly evidenced in Fig. 2a, where larger residuals are seen all over
the profile. The de-speckle procedure produced a sharper profile which
virtually fills the core of the PRF (Fig. 2b). Yet, some systematic
residuals still remain at radius beyond $\sim$10--15 arcsec from the
center, which we assume to be related to an extended emission
component. This component is roughly 10\% of the total emission.

In an attempt to get a better characterization of  the observed profile, a
 combination of an unresolved component represented by the HRI/PSF,
 and an extended component represented by a $\beta$-model are fit to
 the data.  The functional form of the $\beta$-model follows King's
 approximation (1972) for gas confined in an isothermal sphere:
 $S(r)\propto~(1+(r/r_c)^2)^{-3\beta+1/2}$, with S(r) the surface
 brightness a radius r, $r_c$ the core radius and $\beta$ the slope
parameter. Given the uncertainty inherent  with this  approximation
for the extended component,  a simple addition of the PSF and  the
 beta model is fit to the data.

Figure 2c shows the resulting fit corresponding to the composite model:
PRF plus a $\beta$ model.  Fig 2c shows a clear fit improvement to the
observed profile as the systematic residual trend seen in Fig 2b is
now removed; yet, the composite model still overpredicts the observed
emission at radius beyond 10 arcsec. Overall,  the 
$\beta$-model does not provide a statistically acceptable fit with
reduced $\chi^2$ exceeding unity. The main reason for that may reside in 
 the validity of the
beta-model for the specific case of the 
extended gas emission in 3C382. Besides,  there is the fact that the
dominant contribution of the emission is  the unresolved
component,  which makes the modeling of the extended diffuse emission
 difficult. Also, there is the rather  asymmetric morphology of the
emission  which may clearly depart from the simple model used
here. 

Nevertheless,  the fit results from the composite model leaded to a
tight range for  both the core radius and the $\beta$ value. 
The minimum reduced $\chi^2$,  $<1.5-2>$, was obtained
for a core radius $r_c\sim$ 20-30 arcsec (30 - 50 kpc at the distance
of 3C 382) and $\beta\sim 0.7-0.8$. A much simpler composite model
represented by the PRF plus a  Gaussian model for the extended component,
yielded FWHM for the extended component in the 30 - 40 arcsec range.

\subsection{Re-evaluating the PSPC spectrum}

A  single power-law model with spectral index -1.2 provides a fair fit
of the PSPC spectrum of 3C382, with the derived N(H) in good agreement
with the Galactic value.  However, the detection of an  extended
component in the HRI data 
along with the  reported evidence for soft excess emission 
in this source from EXOSAT and ASCA data besides PSPC
(cf. sect. 1) prompt   to a re-evaluation  of the PSPC data.
Because the extended nature of the new component, its  association  with
thermal emission arises as the  most natural explanation.

Due to the short energy range and relatively low spectral
resolution of the PSPC, a complex analysis of the PSPC 
spectrum involving a combination of several models is difficult.
However, the 
evidence for extended emission and the fact that 
the  {\it hard
X-ray spectral index of this source remains constant about the 0.7
value} are additional inputs
that can be used for forcing the PSPC fit with a more complex 
modeling.  

Accordingly, a combination of a
power-law,  with  spectral index fixed at $\alpha$= --0.7, and a
thermal model are fitted to the PSPC  data. In this composite model, $N$(H)
is  fixed
at the Galactic value following Prieto's results (1996). Thus,  the
only parameters that vary freely are the normalizations of the respective
thermal and power-law components and the  temperature of the
thermal component.

The new composite fit (Fig. \pspc) provides a fair representation of
the PSPC spectrum, with a reduced $\chi^2\sim$1.2 and  
constrained values
for all the free parameters. A gas temperature of $0.6^{+0.4}_{-0.1}$ keV is
derived.  
Alternative fits, letting free the index of the power-law
spectrum or  the N(H), lead to larger $\chi^2$
and unconstrained fit parameters.
Fitting a single bremsstrahlung model produces also  an acceptable fit
 but the derived N(H) value is found  lower than
the Galactic value.

The results from the composite model can be compared with those
derived from the single power-law model and from the
single thermal model in Table 1. 
Given the PSPC spectral  limitations together with  the
additional complexity of the composite  model, the derived fit
values may  be subjected to larger uncertainties. In this sense, the
temperature could be largely affected whereas 
integrated fluxes are the least dependent on the adopted
model. The total flux  contribution from the thermal and power-law
components in the composite model
compares with  that    derived from the single
power-law model. The reduced
$\chi^2$ is slighter better in the case of a simple power-law model;
but we consider  the difference   marginal.
The lowest reduced $\chi^2$ is obtained with the thermal model; yet the
derived N(H) in this case  is inferior  than the Galactic value.

\subsection{How do the PSPC fluxes compare with those derived from the 
HRI data?}

Fluxes estimate from the HRI data for the extended and unresolved
component  are primarily 
derived on the basis of   HRI spatial analysis (\S 3).
The total number of counts within the unresolved 
and extended component leads  to 
count rates of $\sim$ 1 cts s$^{-1}$  and  about 0.1  cts s$^{-1}$ 
respectively.  

The lack of spectral resolution of the HRI hampers 
any direct spectral modeling of the X-ray data. Yet, the
availability of the PSPC data allows us to use the same model and fit
parameters as derived from  the PSPC fit (\S 3.1) 
to  convert the HRI  counts to fluxes. These are given in Table 1

The unresolved HRI   component, modeled with a power-law index $\alpha$=
--0.7, yields   a flux  about a 
factor 2 larger than that measured by the PSPC. 
 3C~382 is known to be a
variable source, with reported   maximum-to-minimum variations of up to 
120\% as measured by {\em EXOSAT} (Ghosh \&  Soundararajaperumal 1992). 
Regarding the  {\em ROSAT} observations,  previous HRI  observations of 
the source  in  1992 March
revealed a drop  in the total number of counts by a factor 1.8 with respect 
to the values measured in
the present  observations taken 5 years later. This drop is thus   
 compatible with  the still  lower flux  measured by the PSPC
in 1990.  Thus,  the  difference found between 
the  HRI and PSPC fluxes for the unresolved component appears
 compatible with the observed variability level of the source.

Regarding the extended component,  
bremsstrahlung models with temperatures within the  range   0.6-3 keV 
lead to  corresponding  HRI fluxes 
$ 10 - 6. \times 10^{-12}$ erg~cm$^{-2}~s^{-1}$ respectively, 
i.e., a factor 2 to 3.5 
smaller than that derived from the PSPC for the same component. 
We note  however that as neither the Gaussian approximation nor the $\beta$-model  
provide a statically acceptable  fit 
to that component, the corresponding number of  HRI counts for the extended
component was not derived from the model but from 
the difference between the total 
integrated number of counts  and that integrated 
within the unresolved component represented by the PRF. 
Given the simplicity of the 
method, in particular taking into account that  the 
dominant contribution of  the total  emission 
is the  unresolved component,  
the derived PSPC and HRI fluxes can be  
considered consistent with each other within the  order of magnitude.
Besides,  it is also possible that the emission is  more extended than
what is
detected --at least beyond the radio structure--, but  the surface
brightness there is too low for being detected with the HRI. 

To summarize, the inferred luminosities for  the 
extended and unresolved components 
estimated from  the HRI data  appear compatible in order of magnitude
 with the respective thermal and 
power-law luminosities derived from  PSPC spectrum of the source.

\section{Discussion}

Extended X-ray emission associated with the very  bright, nearby 
radio-loud galaxy 3C~382
is detected.  The analysis of the ROSAT/HRI data shows that about 10\%
of the total 0.2 -2.4 keV emission is compatible with the presence of
an extended component. Assuming that component to be due to hot gas
emitting via bremsstrahlung, and allowing for the contribution of such
thermal component into the PSPC spectrum, it is found that the
non-thermal component of 3C~382 emission becomes consistent with the
extrapolation of the ``well established'' 3C~382 high-energy power-law
spectrum ---above 2 keV--- into the soft X-ray regime.

The temperature of the gas component as formally  derived from the
PSPC fit is 
$0.6^{+0.4}_{-0.1}$ keV. This low temperature contrasts with the high
luminosity of the gas component, $\sim3\times 10^{44} erg~sec^{-1}$.
Taken together the spectral limitation of the PSPC and the complexity
of the model fit, the uncertainty in the temperature could however be
larger.  There is furthermore the possibility that a temperature
gradient dominates the gas emission --- a cooling flow process. In
this case, the PSPC spectrum may be mostly sampling the central gas
region, where the coolest and more dense gas is located.  There are
however other factors that could also have lead to an overestimation
of the gas luminosity. The analysis by Markevitch (1998) on clusters
with strong cooling flows indicates moderate temperature increase of up
to 20\% but luminosity decrease of up to 40\% for the cluster gas
after excising the cooling flow regions.  On the other hand, if
cooling by metals is considered -- for sake of simplicity, pure
bremsstrahlung has been assumed -- the luminosity of the gas could
decrease by about 40\%, assuming a Raymond-Smith model with metal
abundance Z=0.35.

Still, the derived thermal 
luminosity in 3C382 is about two order of magnitude larger
  than that found in isolated, normal elliptical galaxies (Canizares,
  Fabbiano \& Trincheri 1987), and in low-power radio galaxies (Worral
  \& Birkinshaw, 1994) but it is in the range found in powerful radio
  sources (Worrall et al. 1994; O'Dea et al. 1996; Hardcastle et
  al. 1999; Crawford et al. 1999).  Also, the estimated
  core radius for 3C382, $\sim$50 kpc, is  within the range found
 by Crawford et al. and Hardcastle et
 al. in their respective samples of 3CRR radio-loud sources.

Large X-ray halos are often seen in FRI sources, those being
associated with the cluster environment in which they often reside
(e.g. M~87, Perseus, 3C465). These halos largely dominate the ROSAT
emission from these sources.  Besides the outstanding case of Cygnus
A, evidence for clustering is less obvious in classical double radio
sources, particularly at low redshift (cf.  Hill \& Lilly 1991; Miller
et al. 1999).   Unambiguous
extended X-ray emission in powerful FRII  radio galaxies and quasars 
 has mostly being found in sources with  redshift larger
than 0.1 (Hardcastle and Worall, 1999; Crawford et al 1999; O'Dea et
al. 1996); yet, a few low-redshift FRII are reported  to   show extended
X-ray emission (cf. Hardcastle and Worral). In most of these cases,  
the large X-ray  luminosities are  found compatible with thermal
  emission from a moderately rich cluster environment.
 
Comparing with Cyg A,   the archetypal double radio source at z= 0.0574,
3C382 is also one of the few very bright  double sources at low
redshift with extended X-ray emission. Contrarily to  Cyg A which
presents   an optical  narrow  line 
spectrum, 3C382 presents an extreme, in width and
strength, broad permitted line spectrum. If this difference is
 interpreted as due to obscuration of the AGN region in Cyg A, it
 may explain why the
dominant X-ray feature in  Cyg A is emission from a hot diffuse gas
--the AGN component is obscured at X-ray waves-- whereas in 3C382, 
the unresolved X-ray nuclear component --presumably 
associated with the AGN-- dominates the  total X-ray emission, making
more difficult the detection of any extended gas  component.

The  X-ray luminosity of 3C382, of about
$10^{44}erg~s^{-1}$, compares with that of rich Abell clusters (this
is also the case of  Cyg A; yet Cyg A is at the center of a poor
cluster of galaxies).
Longair \& Seldner (1979) derived however a rather poor environment in
the vicinity of 3C~382 on the basis of their cross-correlation
analysis between the radio position and galaxy counts. 
HST/WFPC2 images of 3C382 collected
in parallel mode show an elliptical galaxy with a very bright
unresolved nucleus and a halo very smooth (Martel et al. 1999). 
Yet, within the 2.5
arcminutes field of view (Fig 1b), several small galaxies can easily be
distinguished in the  300 seconds exposure; the WFPC2 images
also show a bright  galaxy at 85 arcsec
Northeast from 3C382,  presenting two at least
extended gaseosus tails of material in the direction of 3C382; two additional
difuse regions located close to 3C382  and in the  direction of the
bright galaxy are also
 apparent. Judging from the HST images, 
3C382 may be residing in a relatively poor cluster environment; 
also, it may be in interaction with that gas-rich  galaxy companion.
Such interaction could have brought plenty
of gas into 3C382.

The luminosity of the halo component in 3C382 would imply a
large mass of gas, of about $10^{11}$Mo, assuming it concentrated in a
sphere of about 50 kpc --the estimate core radius derived from the HRI
spatial analysis-- and a temperature in the 0.6 -1 keV range. An
alternative to the  cluster environment  is that 3C382
it  may  consist of a self-contained gravitational potential deep
enough to restrain such large amount of gas. This could also be the
case of  the radio-louds 3C48 and 3C273,  for which  
extended X-ray
emission is found  but the evidence for a cluster environment 
from optical images is minor  (Crawford et al. 1999).
Evidence for a massive
dark halo in 3C382  comes from the velocity measurements on the
extended ionized gas surrounding this galaxy. Tadhunter et al. (1986) 
detected ionized gas up to 25 kpc from the galaxy
center. The gas follow a  a rotation curve which
extend flat up to those distances with velocities of about 400 km/s
relative to the systemic velocity. Assuming a spherical potential,
the estimated mass within a 25 kpc radius would be $\sim8\times
10^{11}Mo$. This is about the gravitating mass needed to keep the X-ray
gas binded to the galaxy. Following Fornan, Jones and Tucker (1985)
formalism, the total gravitating mass within a 50 kpc radius is
estimated between $8-15 \times10^{11}Mo$ for gas temperatures between 0.6
and 1 keV, which is in the order of magnitude of the mass derived from
the ionized gas kinematics.

The results so far derived show compatible with a cooling flow process
 being dominating the gas emission.  If the extended gas emission is
 modeled as that of a uniform sphere of hot gas emitting via
 bremsstrahlung, for a maximum radius of about 170 kpc (the size of
 the radio structure) and a temperature in the 0.6 - 1 keV range, the
 implied density would be $simeq 6 \times 10^{-3}cm^{-3}$. This yields a
 cooling time of about $4\times 10^{9} yr$, considerable smaller than
 the Hubble time. Thus, a cooling flow process could be operating in
 3C~382.

A better characterization of the extended X-ray emission in 3C382
would demand much larger spatial resolution but also deeper
observations. \\

\bigskip

{\bf Acknowledgments: }

It is a pleasure to thank Peter Predehl for his support regarding the 
de-speckling procedure of the HRI data and G\"unther Hasinger and
Paddy Leahy for critical 
reading of  early versions of the manuscript.

\bigskip
\section*{References}


\ref Black, A. et al. 1992, MNRAS 256, 186

\ref Buehler, P., Courvoisier, T., Staubert, R., Brunner, H., \& Lamer, G.
1995, A\&A, 295, 309

\ref David, L.P. et al. 1998, The ROSAT High Resolution Imager (HRI) Calibration Report: $http://$hea-www.harvard.edu./rosat/rsdc\_www/HRI\_CAL\_REPORT/hri.html

Arnaud, K. A. 1993, ApJ 412, 479

\ref Canizares, C. R., Fabbiano, G. \& Trincheri. G 1987, ApJ 312. 503

\ref Carrili, C. \& Barthel, P. 1996, A\&ARv 7,1
\ref Crawford, C.S. \& Fabian, A.  1995, MNRAS 273, 827

\ref Crawford, C.S., Lehmann, I., Fabian, A., Bremer, M.N. \&
Hasinger, G. 1999, MNRAS 308, 1159

\ref Fanaroff, B. L., \& Riley, J. M. 1974, MNRAS, 167, 31p
 

\ref Ghosh, K. K., \& Soundararajaperumal, S. 1992, ApJ, 389, 179

\ref Hardcastle, M. J. \& Worral, D. 1999, MNRAS 309, 969

\ref Hill G. J., \& Lilly, S. J. 1991, ApJ, 367, 1 


\ref Kaastra J.S, Kunieda \& Awaki H. 1991, A\&A 242, 27

\ref King, I. 1972, ApJ 174, L123
 
\ref Laing, R. A., Riley, J. M., \& Longair, M. S. 1983, MNRAS, 204, 151
\ref Longair, M.S.  \& Seldner, M. 1979, MNRAS 189, 433


\ref Markewitch, M., 1998, ApJ 504, 27

\ref Martel et al., 1999, ApJSS 122, 81

\ref Miller, N.A. et al, 1999, AJ 118, 1988

\ref Mushotzky, R. F., Done, C., \& Pounds, K. A. 1993, ARA\&A, 31,
717

\ref O'Dea, C. P., Worrall, D., Baum, S., \& Stanghellini, C. 1996, AJ, 111, 92

\ref Pounds K.A., Nandra, K., Fink, H., Makino, F. 1994, MNRAS
267, 193

\ref Pounds, K. A., Turner, T. J., \& Warwick, R. S. 1986, MNRAS,
221, 7p
 
\ref Predehl, P. 1998, in preparation

\ref Prieto, M. A. 1996, MNRAS, 282, 421

\ref Ross, R. R., \& Fabian, A. C. 1993, MNRAS, 261, 74
  
\ref Saxton, R. D., Turner, M. J. L., Williams, O. R., Stewart, G. C.,
Ohashi, T., \& Kii, T. 1993, MNRAS, 262, 63

\ref Siebert, J. et al. 1998, MNRAS in press
 
\ref Schmitt, J.H.M.M., G\"udel, M. \& Predehl, P. 1994, A\&A 287, 843
\ref Tadhunter, C., Perez E. \& Fosbury, R. 1986, MNRAS  219,555

\ref Viegas, S. M., \& Contini, M. 1994, ApJ, 428, 113
 

\ref Worral, D. \& Birkinshaw, M. 1994, ApJ 427, 134

\ref Worral, D., Lawrence, C. R., Pearson, T., \& Readhead, A. C. S. 1994, 
ApJ, 420, L17

\ref Wozniak P.R. et al. 1998, MNRAS 299, 449
  
\bigskip
\bigskip

\noindent{\bf Figure captions}

{\bf Figure 1a:}\\
\noindent
HRI contour image  of 3C~382 extracted from channels 1 to 8.  It
is background-subtracted and smoothed with a Gaussian filter with a
FWHM  $\sim$12 arcsec.  Contours are $ 10^{-3} cts~s^{-1}
arcmin^{-2}\times$ (2.8, 4.4, 6, 7.6, 9.2, 10, 16, 40,4000); the first
contour is about the 2$\sigma$ level
measured on the background-subtracted image. 
1 sky-pixel is 0.5 arcsec.\\[2mm]

{\bf Figure 1b:}\\
\noindent
Broad band HST/WFPC2 image of 3C382 (data set u27l6w01-2) at 7200
A an equivalent exposure of 300 seconds. 3C382 is filling the upper
left CCD. The bright companion galaxy (upper right CCD) is North-East
of 3C382. The image covers 2.5 arcminutes.

{\bf Figure 2a:}\\
\noindent
HRI   surface brightness profile of 3C~382 before de-speckling. Points
with error bars (Poissonian noise) are 
the data. Dashed line is the HRI PRF; the continuum line is the  PRF plus
background. The background level is measured in an annulus between 3 to
 5 arcmin from the center (flat part of the profile).
The residuals between the data and the PRF+background are shown below. Note the
large departure from a point source profile due to incorrect
 attitude reconstruction of the ROSAT data (see text).

{\bf Figure 2b:}\\
\noindent
HRI surface brightness  profile of 3C~382 after de-speckling. The interpretation of the figure is as in Fig. 2a.

{\bf Figure 2c:}
\noindent
The same as in Fig 2b but in this case  the  continuum line is the model fit to the surface brightness  profile using a
combination of an unresolved component represented by the HRI PRF (dash)
 and an extended component represented by a $\beta$ model (dots).
 The reduced  
$\chi^2\sim$ of the fit is 1.7-2 over 35 degrees of freedom (d.o.f.).\\[2mm]

{\bf Figure \pspc:}\\
\noindent
Observed and best-model fit to the PSPC spectrum of  3C~382. PSPC data are 
from Prieto (1996). The dotted line 
represents the thermal 0.6-keV component; the dashed line, the power-law 
$\alpha=-0.7$ component. Residuals represent 1-$\sigma$ errors. The
first two  bins corresponding to energies $\simeq0.1 keV$ 
are not included in the fit.

\oddsidemargin 0.01cm
\evensidemargin 0.01cm
\small
\begin{table}[htb]
\begin{center}
\caption{ROSAT results for  3C~382}

\begin{tabular}{lllllcc}\hline \hline

Data &Model & $\alpha$ & KT &  $N$(H) & $\chi^2$/ndf & Flux \\   
 & & & (keV) & $10^{21} {\rm cm}^{-2}$ &- & $ 10^{-11}erg~cm^{-2}$ s$^{-1}$ \\ \hline
PSPC & brems & - & $1.4\pm0.4$ & 0.59$\pm$0.01 & 1.05/29 & $3.7\pm0.5$\\
PSPC & power-law& --1.2$\pm$0.26 & - & 0.79$\pm$ 0.10 & 1.16/29 & $4.5\pm0.2$\\
PSPC &power-law+brems& --0.7 &  $0.65^{+0.4}_{-0.1}$ & 0.78 & 1.23/29
 &$2.5^{+1.5}_{-0.8} + 2^{+1.8}_{-0.7}$\\
HRI& unresolved+extended& --0.7& 0.6 & 0.78 & - & $5.8\pm0.6 + 1.\pm 0.7$\\
\hline \hline
\end{tabular}
\end{center}
Flux is in the 0.2--2.4 keV range and is absorption corrected. Errors
 are 1 sigma correlated errors for one interesting parameter. Errors
 in the HRI fluxes only reflect the uncertainty in the extracted
 number counts to be about  15\% and 50\% for the unresolved and
 extended component respectively. 
The Galactic column density is 0.78 $10^{21}$ cm$^{-2}$. The PSPC count-rate is
 $\sim 2~ cts~s^{-1}$ (total PSPC counts= 1743), 
for the HRI is $\sim 1~cts~s^{-1}$. Single
 power-law model results are from Prieto (1996).

\end{table}

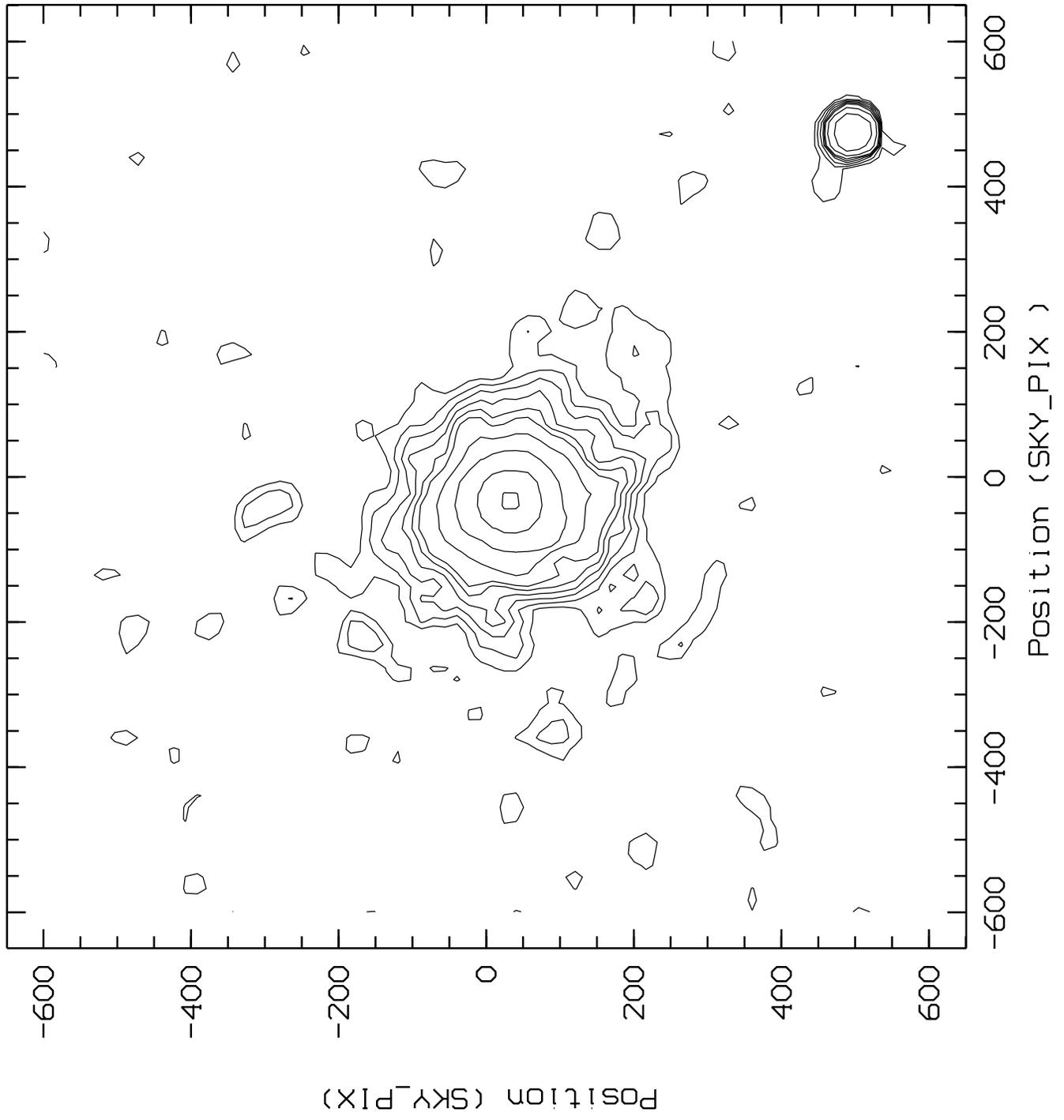
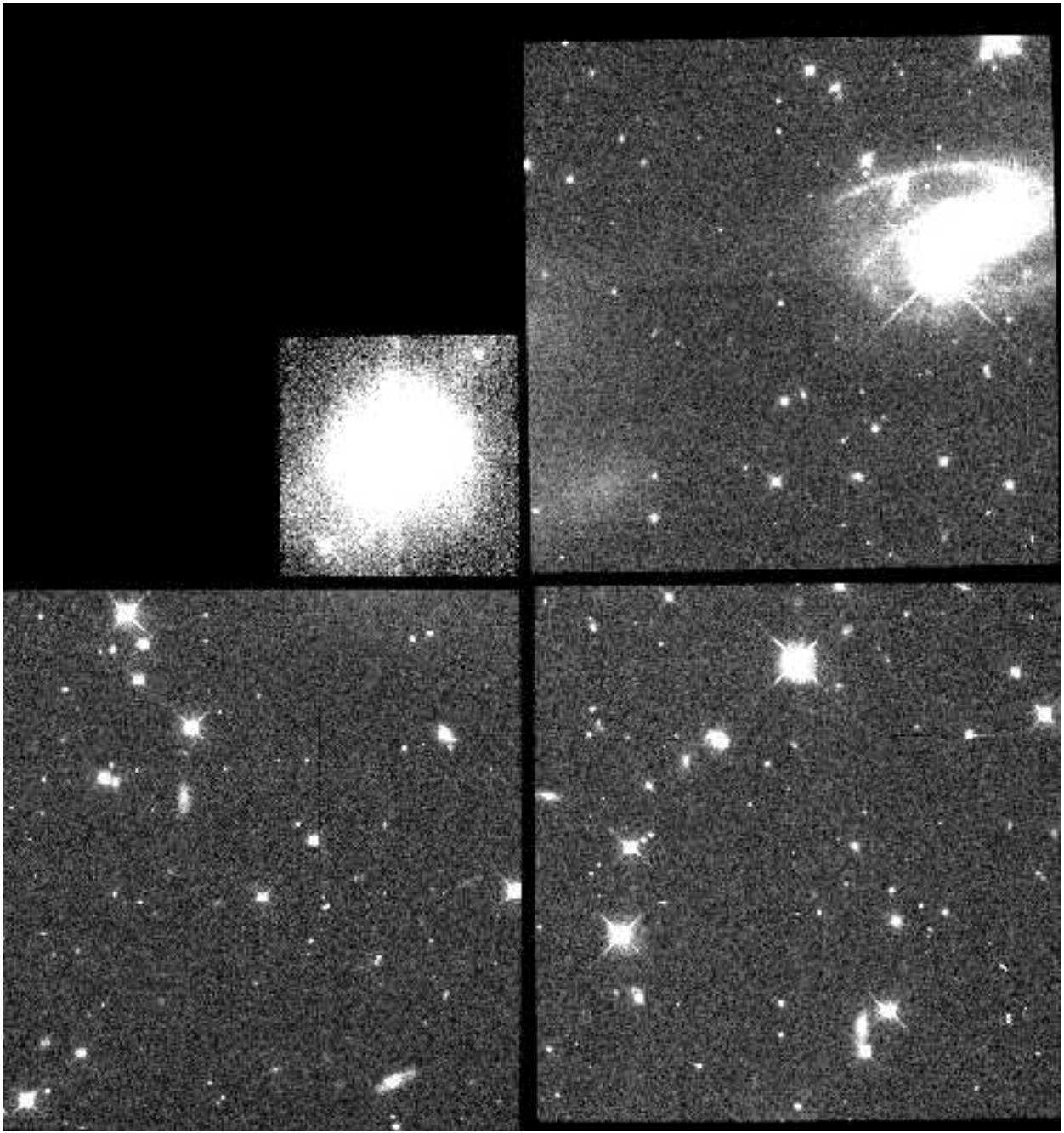
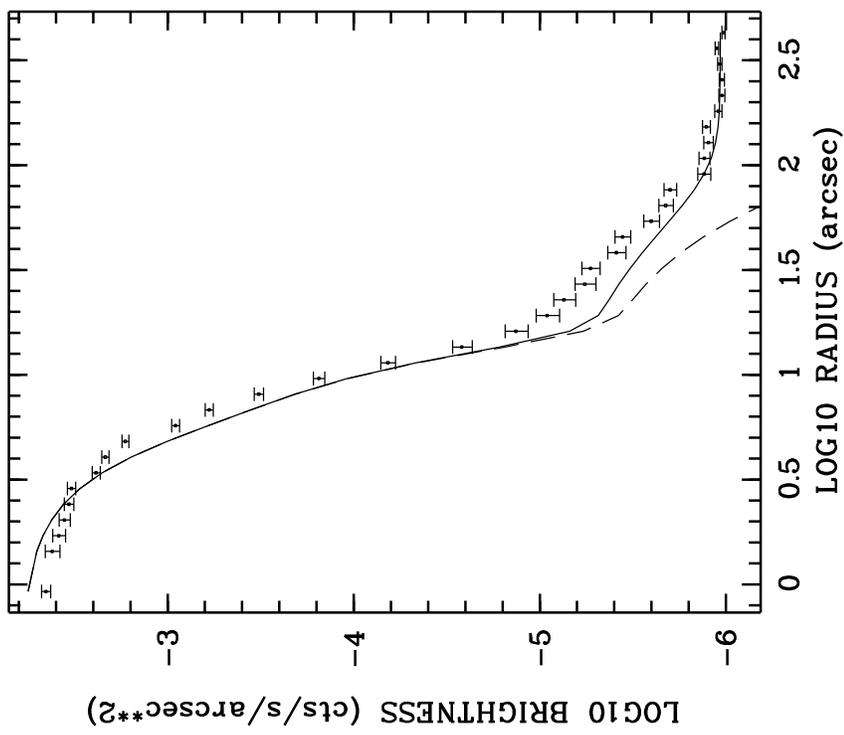
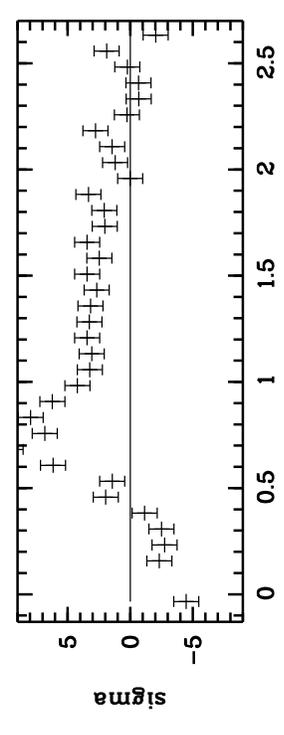
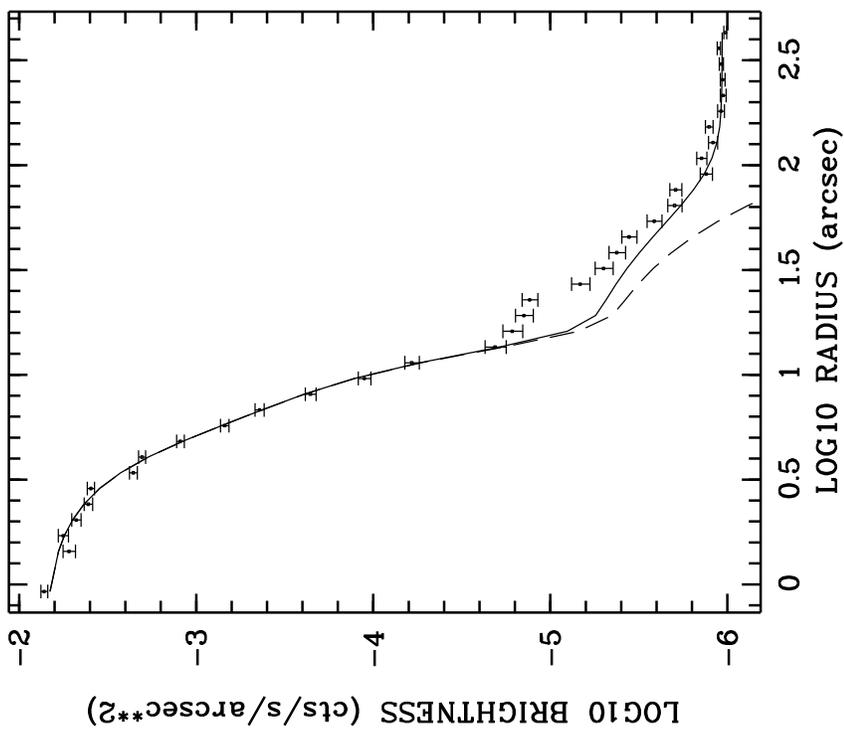
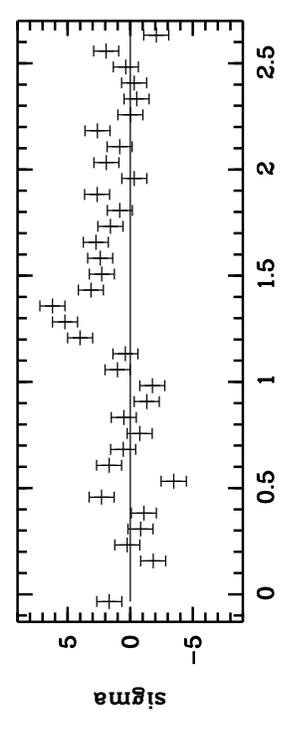
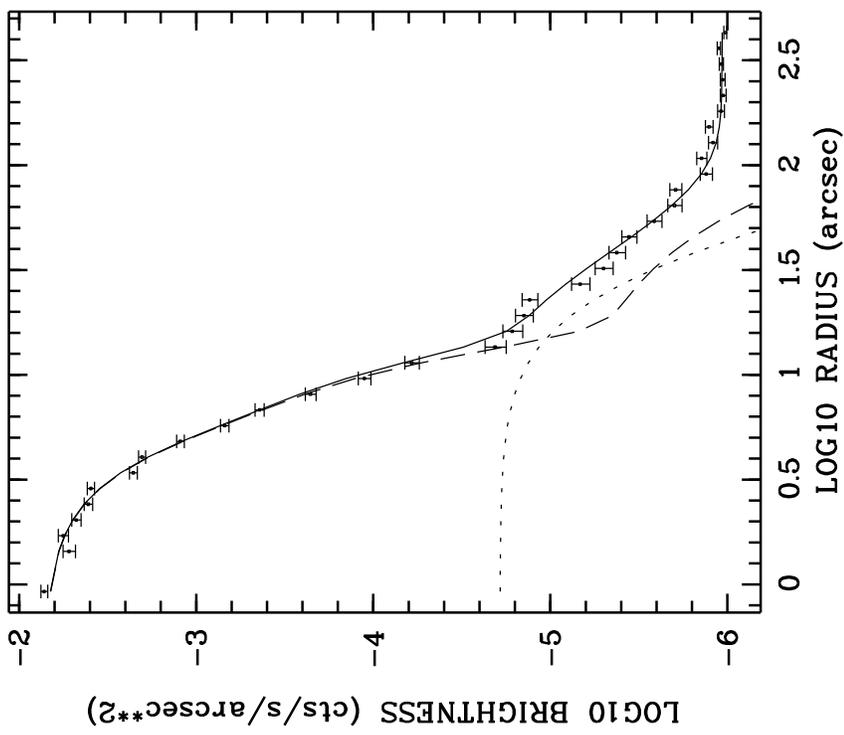
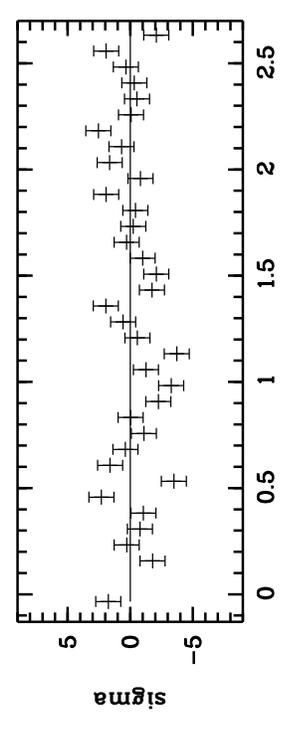
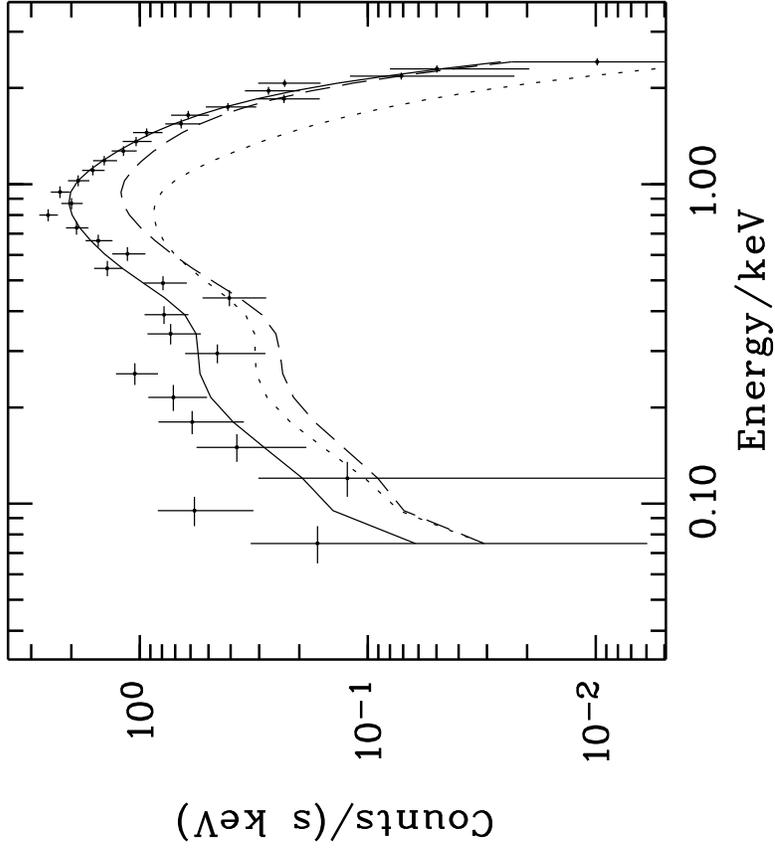
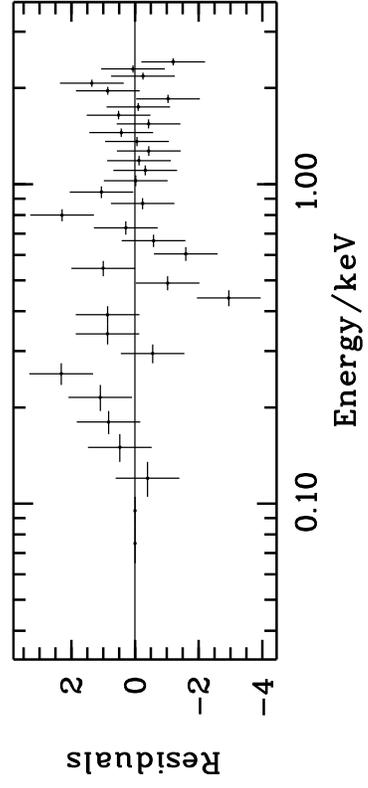
\end{document}